\documentclass[11pt,preprint,usenatbib]{aastex}

\usepackage{amssymb}
\usepackage{rotating}
\usepackage{longtable}
\usepackage{lscape}

\newcommand{\HI}{H\,{\sc i}}
\newcommand{\HII}{H\,{\sc ii}}
\newcommand{\kms}{km\,s$^{-1}$}

\newcommand{\msun}{M$_{\odot}$}

\newcommand{\mhi}{M$_{\rm HI}$}

\newcommand{\Ha}{H$\alpha$}

\newcommand{\HIIbold}{{H\footnotesize{\bf{II}}}}
\newcommand{\nhi}{\mbox{$N_{\rm HI}$}}
\newcommand{\Lya}{\mbox{Ly$\alpha$}}
\newcommand{\cm}{cm$^{-2}$}

\newcommand{\oiii}{\hbox{[O\,{\sc iii}]}}

\newcommand{\nii}{\hbox{[N\,{\sc ii]}}}
\newcommand{\mhilb}{\mbox{$M_{\rm HI}$-to-$L_B$}}

\slugcomment{accepted by the Astronomical Journal}

\shorttitle{Intergalactic \HII\ Regions}
\shortauthors{Ryan-Weber et al.}

\begin{document}

\title{Intergalactic \HIIbold\ Regions Discovered in SINGG}

\author{E. V. Ryan-Weber\altaffilmark{1,2},
G. R. Meurer\altaffilmark{3},
K. C. Freeman\altaffilmark{4},
M. E. Putman\altaffilmark{5},
R. L. Webster\altaffilmark{1},
M. J. Drinkwater\altaffilmark{6},
H. C. Ferguson\altaffilmark{7},
D.   Hanish\altaffilmark{3},
T. M. Heckman\altaffilmark{3},
R. C. Kennicutt, Jr.\altaffilmark{8},
V. A. Kilborn\altaffilmark{9},
P. M. Knezek\altaffilmark{10},
B. S. Koribalski\altaffilmark{2},
M. J. Meyer\altaffilmark{1},
M. S. Oey\altaffilmark{11},
R. C. Smith\altaffilmark{12},
L.   Staveley-Smith\altaffilmark{2},
and M. A. Zwaan\altaffilmark{1}
}

\altaffiltext{1}{School of Physics, University of Melbourne,  
                 VIC~3010, Australia.}
\altaffiltext{2}{Australia Telescope National Facility, CSIRO, 
                 P.O. Box 76, Epping, NSW~1710, Australia.}
\altaffiltext{3}{Department of Physics and Astronomy, The Johns Hopkins
                 University, Baltimore, MD 21218-2686, USA}
\altaffiltext{4}{Research School of Astronomy \& Astrophysics, Mount Stromlo
                 Observatory, Cotter Road, Weston, ACT~2611, Australia.}
\altaffiltext{5}{CASA, University of Colorado, Boulder, CO 80309-0389, USA.}
\altaffiltext{6}{Department of Physics, University of Queensland,
                 Brisbane, Queensland, 4072, Australia}
\altaffiltext{7}{Space Telescope Science Institute, 3700 San Martin
                 Drive, Baltimore, MD 21218, USA}
\altaffiltext{8}{Steward Observatory, University of Arizona, Tuscon,
                 AZ 85721, USA}
\altaffiltext{9}{Jodrell Bank Observatory, University of Manchester, 
                 Macclesfield, Cheshire, SK11 9DL, U.K.}
\altaffiltext{10}{WIYN Observatory, 950 N. Cherry Avenue, Tuscon, AZ
                 85719, USA}
\altaffiltext{11}{Lowell Observatory, 1400 West Mars Hill Rd, USA}
\altaffiltext{12}{Cerro Tololo Inter-American Observatory, Casilla
                 603, Chile}

\begin{abstract}
A number of very small isolated \HII\ regions have been discovered at
projected distances up to 30 kpc from their nearest galaxy. These
\HII\ regions appear as tiny emission line objects in narrow band
images obtained by the NOAO Survey for Ionization in Neutral Gas
Galaxies (SINGG). We present spectroscopic confirmation of four isolated
\HII\ regions in two systems, both systems have tidal \HI\ features. The
results are consistent with stars forming in interactive debris due to
cloud-cloud collisions. The \Ha\ luminosities of the isolated
\HII\ regions are equivalent to the ionizing flux of only a few O
stars each. They are most likely ionized by
stars formed in situ, and represent atypical star formation in the low
density environment of the outer parts of galaxies. A small but finite
intergalactic star formation rate will enrich and ionize the
surrounding medium. In one system, NGC 1533, we calculate a star
formation rate of $1.5\times10^{-3}$ \msun yr$^{-1}$, resulting in a
metal enrichment of $\sim1\times10^{-3}$ solar for the continuous
formation of stars. Such systems may have been more common in the past
and a similar enrichment level is measured for the `metallicity floor'
in damped Lyman-$\alpha$ absorption systems.

\end{abstract}


\section{Introduction}
\label{sect:intro}

\HII\ regions signifying the presence of highly ionizing OB stars are
usually found in the luminous inner regions of galaxies
\citep[e.g.][]{Martin01}. \HII\ regions are also located in the faint
outer arms of spirals \citep[e.g.][]{Ferguson98b}, and as single or
multiple star forming knots in narrow emission line dwarfs (\HII\
galaxies). In each case, new stars are formed in the vicinity of an
existing stellar population. However, recent observations by
\cite{Gerhard02} have spectroscopically confirmed an isolated compact
\HII\ region on the extreme outskirts of a galaxy (NGC 4388) in the
Virgo cluster. Several luminous \Ha-emitting knots have also been
discovered in a compact group in the A1367 cluster \citep{Sakai02}. In
these cases it appears that the \HII\ regions are due to newly formed
stars where no stars existed previously, albeit in a galaxy cluster
environment.

As they evolve, OB stars increase the metal abundance in their local
environment. Absorption line studies show that the intergalactic
medium (IGM) and galaxy halos, including our own, are enriched
\citep[e.g.][]{Chen01,Tripp02,Collins03}. Isolated \HII\ regions
provide a potential source for this enrichment. In situ star formation
in the IGM offers an alternative to galactic wind models to explain
metal enrichment hundreds of kilo-parsecs from the nearest galaxy.

Here we present a number of very small isolated \HII\ regions that
have been discovered by their \Ha\ emission in the narrow band images
obtained by the NOAO Survey for Ionization in Neutral Gas Galaxies
(SINGG). SINGG is an \Ha\ survey of an \HI-selected sample of nearby
galaxies. The survey is composed of nearly 500 galaxies from the \HI\
Parkes All-Sky Survey (HIPASS, Barnes et al.  \citeyear{Barnes01};
Meyer et al.  \citeyear{Meyer03}), of these about 300 have been
observed in \Ha.  Since a gaseous reservoir is a prerequisite for star
formation, SINGG measures a broad census of star formation in the
local Universe.  The \HII\ regions appear as tiny emission line
objects at projected distances up to 30 kpc from the apparent host
galaxy. \HII\ regions are defined as isolated when they are projected
at least twice the $\mu_R =$ 25 ${\rm mag\, arcsec^{-2}}$ isophotal
radius from the apparent host galaxy. This is typically much further
than outer disk \HII\ regions in spiral galaxies \citep{Ferguson98b}.
In fact, for the systems discussed in detail here it is not totally
clear whether the isolated \HII\ regions are even bound to the
apparent host, hence we refer to them as ``intergalactic''.  Their
high equivalent widths suggest they are due to newly formed stars
where no stars existed previously.

In Section~2 spectra is presented for five isolated \HII\ regions
candidates in three systems. These five candiates are referred to as
the spectroscopically {\it{detected}} emission line objects or \HII\
region candidates. All but one source has \Ha\ detected at a
comparable recessional velocity to the nearest galaxy, two sources are
also detected in \oiii\ as further confirmation.  These four objects
are referred to as the spectroscopically {\it{confirmed}} isolated
\HII\ regions. Optical spectra and \HI\ distributions for all three
systems are described in Sections~3$-$5. In Section~6 models of the
underlying stellar population, scenarios for star formation,
enrichment of the IGM, implications of the intergalactic star
formation rate, and the possibility that isolated \HII\ regions are
progenitors of tidal dwarf galaxies are discussed. H$_0$=75 \kms
Mpc$^{-1}$ is used throughout.

\section{Observations}
\label{sect:obs}

Continuum R-band and narrow band \Ha\ images of local gas-rich galaxies
were taken with the CTIO 1.5m telescope as part of SINGG.  At least
three images were taken in each band, with small ($\lesssim 1'$) dithers
between images.  Each image was processed through overscan and bias
subtraction, followed by division by a flatfield derived from both dome
and twilight sky flats.  The images in each band were aligned and
combined.  The R or narrow-band image with the better seeing was
convolved to match the poorer seeing of the other image, and then the R
image was scaled and subtracted from the narrow band image to produce a
net \Ha\ image.  The final images have a pixel scale of 0.43$''$ per
pixel and subtend a field of view of nearly 15$'$.  The properties of
the SINGG images are given in Table~\ref{table:images}.  The Table
includes the $5\sigma$ point source detection limit and the large scale
surface brightness limit for both R \& \Ha\ images.  The latter was
determined from the rms variation in the mean background level
determined in boxes 35 pixels on a side after iteratively clipping
pixels that deviate by more than 5 times the pixel to pixel rms from
each box.  The box to box variation of the mean is less than 1\%\ of the
sky level in all cases.

The candidate isolated \HII\ regions were identified as unresolved
high equivalent width (EW) sources outside the optical disk of each
galaxy. Individual exposures were checked and candidates with sharp
edges (likely due to cosmic ray residuals) or bright continua were
rejected. Aperture photometry was performed on each of the isolated
\HII\ regions in the flux calibrated SINGG images using
{\sl{idlastro}} routines in IDL yielding \Ha\ and R fluxes.  Since the
\HII\ regions have high equivalent widths, the R count rates were
corrected for \Ha\ line emission using the formula
\begin{equation}
C_{\rm R}' = C_{\rm R} - \frac{T_{\rm R}(\lambda)}{T_{\rm NB}(\lambda)}*C_{\rm
  H\alpha} 
\end{equation}
where $C_{\rm R}'$ is the corrected R count rate, $C_{\rm R}$ and
$C_{\rm H\alpha}$ are the measured count rates in the R and net \Ha\
image, and $T_{\rm R}(\lambda)$ and $T_{\rm NB}(\lambda)$ are the
throughputs of the filter evaluated at the wavelength $\lambda$ of
\Ha.  For the spectroscopically detected cases, $\lambda$ was taken
from the spectra; otherwise we assumed that SINGG was detecting net
\Ha\ emission redshifted by the \HI\ heliocentric radial velocity.
The results are given in Table~\ref{table:eldots2}. The
spectroscopically detected isolated \HII\ regions have \Ha\ fluxes in
the range 6.9 to 11$\times10^{-16}$ erg s$^{-1}$\cm. Assuming the
distance to each isolated \HII\ regions is the same as the host galaxy
in each system, the \Ha\ luminosities are $4\pi D^2f_{H\alpha}=
3.5\times10^{36}$ to $3.5\times10^{38}$ erg s$^{-1}$. In most cases
the isolated \HII\ regions are barely detected in continuum emission
in the SINGG R images with a typical $5\sigma$ detection limit of
around $1\times10^{-18}$ erg s$^{-1}$\cm\AA$^{-1}$ (see
Tables~\ref{table:images} and \ref{table:eldots2}). Isophotal radii at
$\mu_R = 25 {\rm mag\, arcsec^{-2}}$ were measured for each galaxy
using the {\sl{ellipse}} task in IRAF. The galaxy-\HII\ region
separations are given as a function of these radii in
Table~\ref{table:eldots1}. 

\begin{figure} 
\vspace{20cm}
\includegraphics{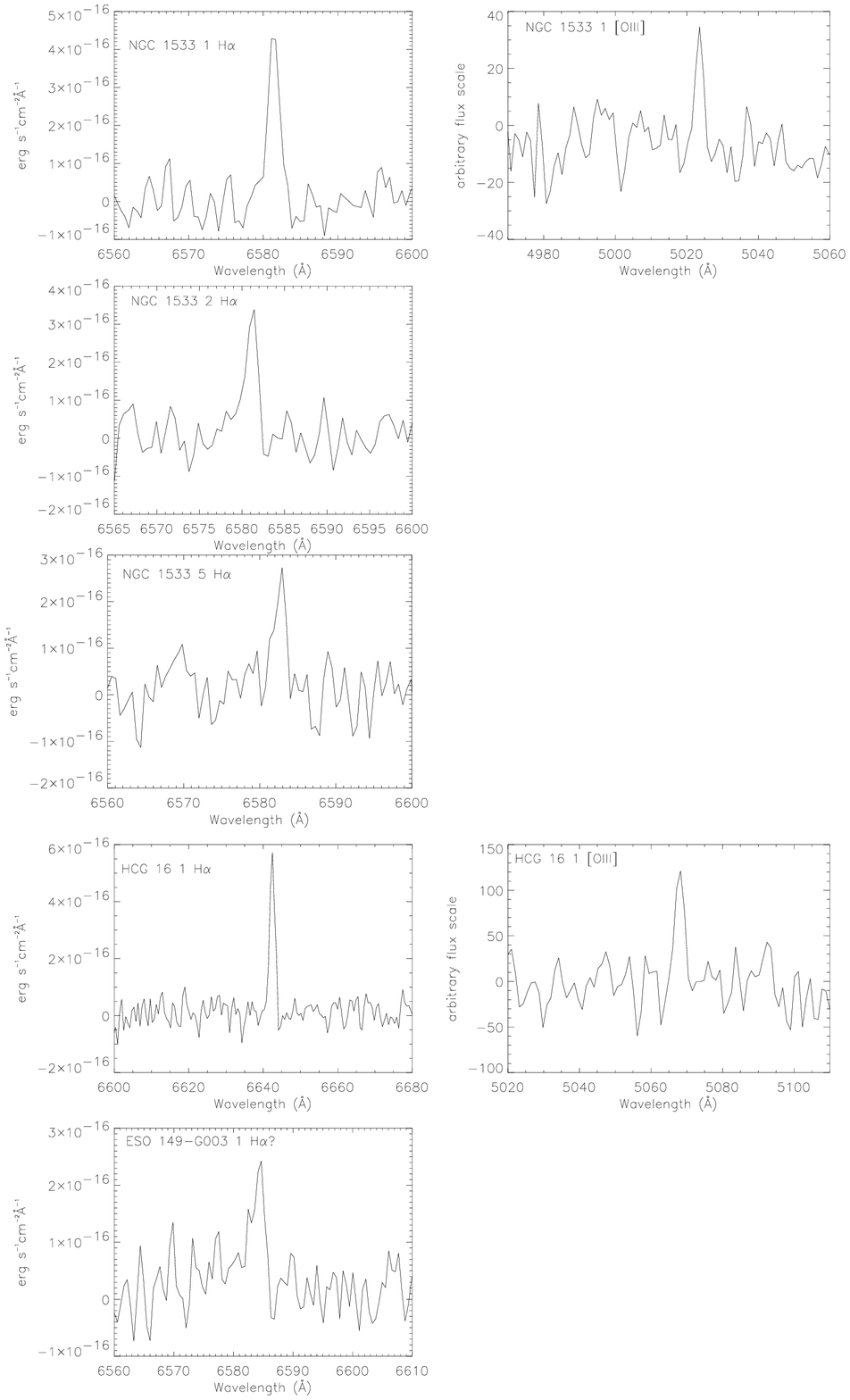} 
\caption{Emission lines from detected isolated \HII\ region
candidates: \Ha\ on the left and \oiii\ on the right. The \Ha\ spectra
have been normalised by the integrated \Ha\ flux from the SINGG images
given in Table~\ref{table:eldots2}, corrections for an \nii\ component
have not been made.}.
\label{fig:eldotspec}
\end{figure}

Spectra of 11 isolated \HII\ region candidates were obtained with the
double beam spectrograph (DBS) on the RSAA 2.3m telescope in September
2002. Due to the very low continuum flux of the emission line objects,
it was necessary to use a nearby star to align the slit on each \HII\
region candidate in the DBS auto-guider. Three 2000 second exposures
were taken of each object. The spectra were reduced using standard
procedures in IRAF. The dispersion of the red spectra (6000 - 7000
\AA) was 0.55 \AA\ per pixel, corresponding to a resolution of 1.1
\AA\ per pixel or 50 \kms. The dispersion of the blue spectra (3500 -
5400 \AA) was 1.1 \AA\ per pixel, corresponding to a resolution of 2.2
\AA\ per pixel or 150 \kms. An additional two exposures of the
candidate \HII\ region near ESO 149-G003 were obtained in October 2003
using the same DBS set-up. These additional exposures have been
combined with the first spectra to give the result in
Figure~\ref{fig:eldotspec}.

Five of the 11 observed isolated \HII\ region candidates were
detected. The non-detected candidates mostly have $F_{\rm H\alpha} <
4\times10^{-16}\, {\rm erg\, cm^{-2}\, s^{-1}}$ (with the exception of
the candidate \HII\ region near NGC 1314) and have a range of
continuum fluxes. In each case the DBS red spectrum fully covers the
narrow band SINGG filter range. Since the spectra do not confirm the
reality of these 6 isolated \HII\ region candidates, they will not be
discussed further. Four of the 5 detected isolated \HII\ regions have
confirmed emission lines within the narrow band filter's passband
close to the expected position of \Ha\ (6563 \AA) at recessional
velocities close to that of their respective host galaxies (measured
from HIPASS, see Table~\ref{table:eldots1}, column 2). The recessional
velocity measured from each \Ha\ line is given in
Table~\ref{table:eldots2}. Some isolated \HII\ regions were also
detected in \oiii\ (5007 \AA), confirming that the emission seen in
the SINGG images is indeed \Ha. The presence of both \Ha\ and \oiii\
lines in these cases places the isolated \HII\ regions at comparable
recessional velocities to the galaxy (or galaxies) in each field, and
rules out the possibility that they are background emission line
systems. Spectra for the 5 detected isolated \HII\ regions are given
in Figure~\ref{fig:eldotspec}. The \Ha\ spectra (red arm) have been
normalised by the continuum subtracted \Ha\ fluxes from the SINGG
images. The \Ha\ line fluxes have not been corrected for \nii\
contamination. For the FWHM$\sim$35\AA\ filters used in this study we
expect the combined \nii\ 6548+6583 contamination to be $<$10\% if the
line flux ratio \nii\ 6548/\Ha = 0.35. The \oiii\ line lies in the
blue arm and the flux calibration of the spectra is uncertain.
Details of the 3 systems, which include the 5 spectroscopically
detected isolated \HII\ regions, are discussed in the next section.

In addition to the \Ha\ images and spectra, Australia Telescope
Compact Array (ATCA) \HI\ maps are available for two systems, NGC 1533
and ESO~149-G003. The ATCA data reduction is detailed in
\cite{Ryan-Weber03b}.  The two datasets have a restored beam of
$68\arcsec\times65\arcsec$ and $79\arcsec\times61\arcsec$
respectively. The velocity resolution is 3.3~\kms\ and the RMS noise
is 3.7 mJy beam$^{-1}$ per channel, corresponding to a 3$\sigma$
column density limit (over a line width of 40~\kms) of
3.2$\times10^{19}$~\cm.

\section{Three Systems with Detected Isolated \HII\ regions}

\subsection{NGC~1533}
\label{sect:ngc1533}

\begin{figure} 
\vspace{9cm}
\includegraphics{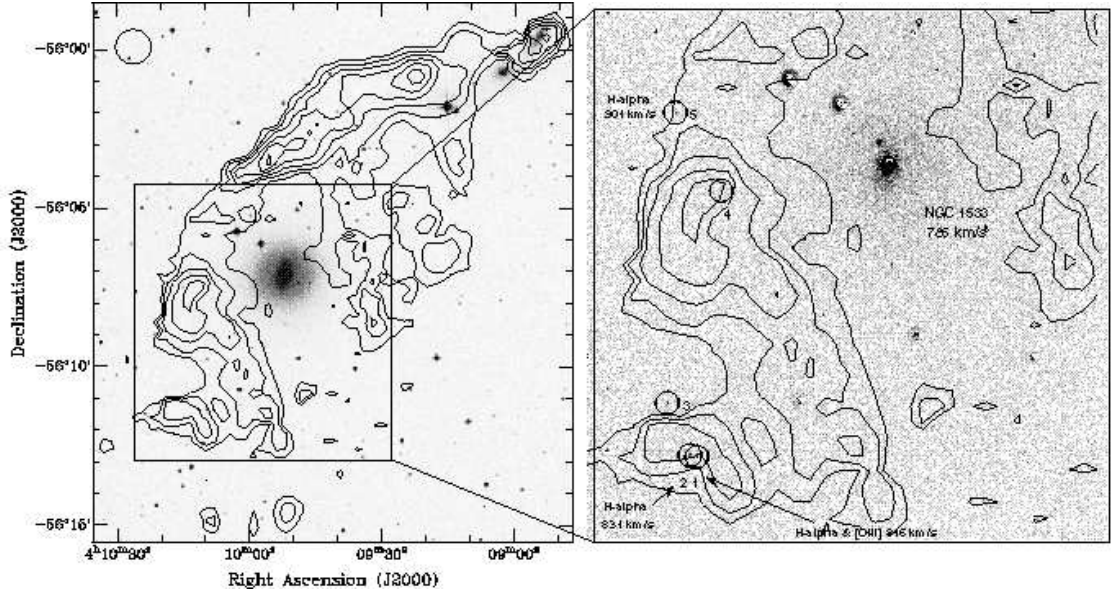} 
\caption{NGC 1533: DSS image with ATCA \HI\ contours overlaid at 1.6,
  2.0, 2.4, 2.8, 3.2, 3.6 and 4 $\times10^{20}$ \cm. The beam is given
  in the top left corner. The insert shows the continuum subtracted
  \Ha\ image with the isolated \HII\ regions labelled and \Ha\
  velocities given, where available.}
\label{fig:ngc1533}
\end{figure}

Figure~\ref{fig:ngc1533} shows a DSS image of NGC~1533 overlaid with
ATCA \HI\ contours. The insert has the same contours overlaid on the
\Ha\ image with the isolated \HII\ regions labelled. The \HI\
distribution around NGC~1533 consists of two major arcs, the NW cloud
and the SE cloud. \HI\ gas with column densities below the lowest
contour close the \HI\ in a ring. No obvious optical counterpart to
this ring is seen in the DSS nor R SINGG image. The total \HI\
mass of the system (based on the total flux density from HIPASS of
67.6 Jy beam$^{-1}$\kms) is 7$\times10^{9}$~\msun. The SE cloud
contains $\sim\frac{1}{3}$ of this total \HI\ mass
(2.4$\times10^{9}$~\msun). The projected distance between the \HI\
arcs and the optical centre of NGC~1533 ranges from 2\arcmin\ to
11.7\arcmin, corresponding to a projected physical length between 12
and 70 kpc.

NGC 1533 is an S0 galaxy located 1$^{\circ}$ from the center of the
Dorado group. The two smaller galaxies in the NW corner of the image
are IC 2039 (closest to NGC 1533, uncertain redshift, no \HI\
detected) \& IC 2038 (contains associated \HI). The peculiar
distribution of \HI\ is thought to arise from the destruction of
a galaxy to form a tidal remnant around NGC 1533. If the \HI\ was
stripped from IC2038/9, these galaxies would need to account for all
the \HI\ in the system and therefore have \mhilb\ ratios greater than
15, which is not very likely for their morphologies. The progenitor is
more likely a low surface brightness (LSB) galaxy with a moderate \HI\
mass, whose optical counterpart is now too diffuse to
identify. N-body/SPH numerical simulations showing the orbital
evolution of a LSB galaxy in NGC~1533's gravitational potential
support this hypothesis \citep{Ryan-Weber03a}.

The velocities of the three confirmed isolated \HII\ regions (1, 2 and
5), 846, 831 and 901 \kms, compare well with the velocity of NGC~1533
at 785 \kms\ and lie within the range of \HI\ velocities in the SE
cloud (883 \kms\ with a width at 50\% peak, $w_{50}=71$
\kms). Interestingly, the isolated \HII\ regions do not appear to be
correlated with the densest regions of \HI\ and are located in the SE
cloud only. At this resolution ($\sim$6 kpc) the densest region of
\HI\ is the central part of the NW cloud. The stellar concentrations of
tidal dwarf galaxies are located in the densest regions of \HI, mapped
in 21-cm at similar resolutions \citep[$\sim$ 4kpc,
  e.g.][]{Duc00}. Furthermore, the \HI\ in the SE cloud has velocity
dispersions up to 30 \kms\ and velocity gradients in the range 7-50
\kms kpc$^{-1}$, making it an unlikely site for star formation. Star
formation usually requires the gas to have a low velocity dispersion
in order to collapse gravitationally.

\subsection{HCG~16}
\label{sect:hcg16}

\begin{figure} 
\vspace{9cm}
\includegraphics{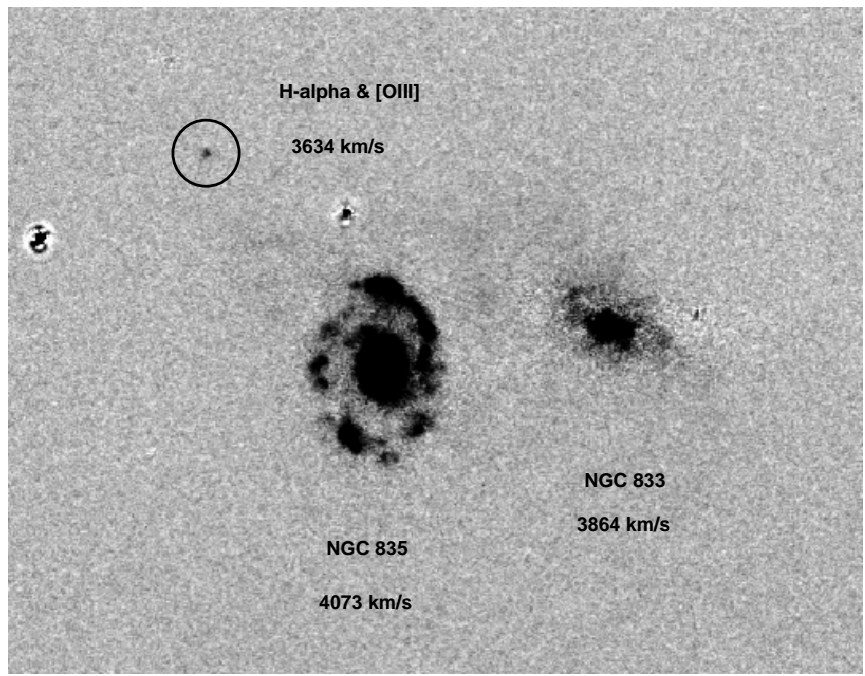} 
\caption{HCG 16: Continuum subtracted \Ha\ image with the isolated
  \HII\ region and two members of the galaxy group labelled. The two
  other objects in the field are residuals of foreground stars.}
\label{fig:hcg16}
\end{figure}

The isolated \HII\ region in the compact group HCG~16, shown in
Figure~\ref{fig:hcg16}, is near the two galaxies NGC~835 (SBab) and
NGC~833 (Sa). The velocity of the isolated \HII\ region (3634 \kms)
sits on the lower edge of the \HI\ emission measured by HIPASS
(velocity = 3917 \kms, $w_{50}=$ 288 \kms, $w_{20}=$ 391 \kms) and
below the optical velocities of NGC 835 and NGC 833 (4073 and 3864
\kms\ respectively, from NASA/IPAC Extragalactic Database, NED). The
2D spectrum shows the \Ha\ emission line from NGC 833 at 3864 \kms\
and diffuse emission decreasing in velocity to $\sim$3700 \kms\ half
way along the line towards the isolated \HII\
region. \cite{Verdes-Montenegro01} have published a VLA map of HCG 16,
showing \HI\ in NGC 835 and 833 with a large tidal feature to the NE
(overlapping the isolated \HII\ region position) that joins other
group members several arcminutes away to the east.

\subsection{ESO~149-G003}
\label{sect:eso149}

\begin{figure} 
\vspace{9cm}
\includegraphics{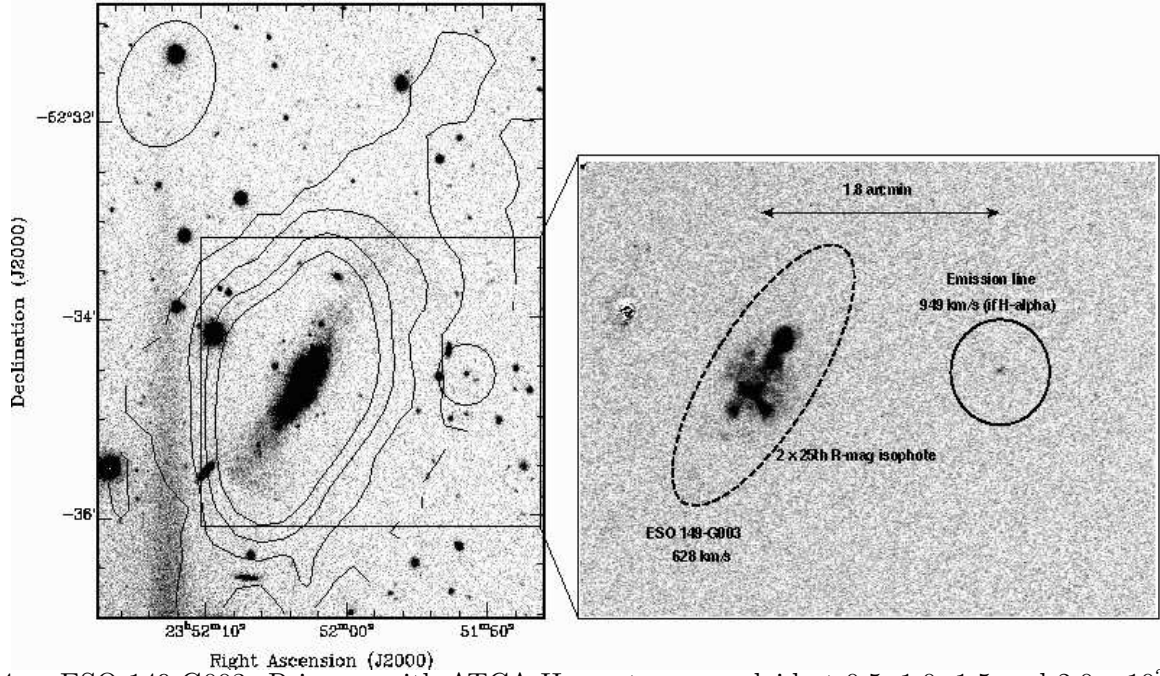} 
\caption{ESO~149-G003: R image with ATCA \HI\ contours overlaid at
  0.5, 1.0, 1.5 and 2.0 $\times10^{20}$ \cm. The beam is given in the
  top left corner. The insert shows the continuum subtracted \Ha\
  image with the isolated \HII\ region candidate and galaxy labelled.}
\label{fig:eso149}
\end{figure}

The velocity of the isolated \HII\ region candidate near the irregular
galaxy ESO~149-G003 (949 \kms) is quite offset from its apparent host
galaxy (\HI\ velocity of 576 \kms\ and $w_{50}=39$ \kms). The
long-slit spectrum of the \HII\ region candidate also includes the
southern part of the galaxy, and the galaxy \Ha\ emission line has a
measured velocity of 628 $\pm50$ \kms. This 52 \kms\ deviation, just
inside the quoted uncertainty, could be due to the slit being aligned
along the southern part of the galaxy only, although the \HI\ spectrum
does not show a large velocity gradient.  A velocity difference of 321
\kms\ is measured between the galaxy and \HII\ region candidate optical
emission lines, assuming the single emission line is indeed \Ha.  The
narrow \HI\ profile shows no anomalous velocity gas. Follow-up ATCA
observations show no \HI\ emission at the velocity and position of the
emission line object to \mhi\ $\leq 8\times 10^5$ ($3\times 10^6$)
\msun\ at a $3\sigma$ limit, and assuming a distance of 6.5 (12) Mpc.
ESO~149-G003 seems quite isolated, and the nearest galaxy in both NED
and HIPASS (Meyer et al. \citeyear{Meyer03}) is ESO~149-G013 at 1500
\kms, 1.6\degr\ ($\sim$560 kpc) away.  However, ESO~149-G003 does show
signs of a flared or warped optical disk at the edges, suggesting an
interaction has taken place (see Figure~\ref{fig:eso149}). A strong
positive correlation between warping in late-type galaxies and
environment \citep{Reshetnikov98} has been largely attributed to tidal
interactions. However, since only one emission line (\Ha) is detected
in this case, the possibility that this candidate isolated \HII\
region is a background emission line source cannot be ruled out, for
example H$\beta$ at z$\sim$0.4 or \oiii\ at z$\sim$0.3.

\section{Discussion}
\label{sect:discussion}
\subsection{Underlying Stellar Population}

The \Ha\ luminosities of the detected isolated \HII\ regions (
$3.5\times10^{36}$ to $3.5\times10^{38}$ erg s$^{-1}$) place them at
the low luminosity end of the \HII\ region luminosity function
\citep[e.g.][]{Oey98}. The \Ha-luminosity of an \HII\ region is
proportional to the ionizing photon luminosity ($Q_0$) above the Lyman
limit (912 \AA) from nearby stars. From equation 5.23 in
\cite{Osterbrock89}, $Q_0=(\alpha_{B}/
\alpha^{eff}_{H\alpha})\times(L_{H\alpha}/E_{H\alpha})=7.33\times
10^{11} L_{H\alpha}$, and using the ionizing luminosity of an O5V star
\citep{Vacca96}, each isolated \HII\ region is illuminated by the
equivalent of $0.1-8$ O5V stars each. The least luminous detected
isolated \HII\ region can be ionized by a single O9.5 star.

The underlying stellar population is very weak in most of the
spectroscopically detected emission line objects, in two of the five
cases it is undetected in the SINGG images. This makes it difficult to
constrain whether the isolated \HII\ regions are ionized by a single
isolated massive star, or whether the massive star or stars represent
the `tip of the iceberg' of a cluster. A single massive star could
have formed spontaneously; examples of this exist in the disk of the
Milky Way, where IR observations show an isolated massive star
\citep{Ballantyne00}. However, most massive stars form as part of a
cluster \citep{Clarke00}, with a characteristic Initial Mass Function
(IMF). The very low continuum emission does rule out a significant
underlying stellar population and suggests that isolated \HII\ regions
are due to newly formed clusters where no stars existed
previously. The low continuum emission also separates isolated \HII\
regions from \HII\ galaxies and tidal dwarf galaxies. Three of the
four confirmed isolated \HII\ regions have EW(\Ha)$>$1000 \AA. By
comparison, \HII\ regions in the outer arms of spiral galaxies (beyond
the B 25th-magnitude isophote) measured by \cite{Ferguson98a}
have an average EW(\Ha)=364 \AA. Furthermore, most of the group of
star forming dwarf galaxies in A1367 have EW(\Ha)$<$100 \AA\
\citep{Sakai02}.

Upper limit estimates of the underlying stellar population can be
obtained from star formation models, such as Starburst99
\citep{Leitherer99}. The ratio of $Q_0$ to the continuum luminosity,
$L_v$, can be used to find the age of the population. Since very
little continuum emission is detected from the isolated \HII\ regions,
the ratio $Q_0/L_v$ is large, suggesting a very young age, in the
range 3 to 7$\times10^6$ years, for the parameters of the five
detected \HI\ regions. Indeed young ages are expected since we
selected isolated \HII\ region candidates with high equivalent
widths. Using the Starburst99 model with a Salpeter IMF,
$M_{up}=100$\msun, and metallicity of 0.4 solar, a very young
instantaneous burst is predicted to have a photon luminosity in the
range log$Q_0=51.3$ to 52.7 photons sec$^{-1}$ for a system with
10$^6$ \msun. The same model predicts 1.2 to 4.5$\times10^{3}$ O stars
(in the spectral range O3 to O9.5). Scaling $Q_0$ with $Q_0$(\HII\
region) the number of O stars ionizing each isolated \HII\ region is
in the range $4-7$, corresponding to a total cluster mass of
$0.9-1.8\times10^{3}$ \msun. The HCG~16 isolated \HII\ region is
significantly more luminous in \Ha, according to this model, it would
consist of 23 O stars in a cluster of $5\times10^{3}$ \msun.
The ESO~149-G003 isolated \HII\ region, unlike the others, does have
considerable continuum emission in the R image (see
Table~\ref{table:eldots2}). This suggests a slightly older population
(7$\times10^6$ years) with 1.4 O stars and cluster mass of
1.2$\times10^{3}$ \msun. These models account for the nebula emission
in the continuum luminosity.

Continuous star formation models predict stellar population ages in
the range 6$\times10^{6}$ to 1.5$\times10^{7}$ years for isolated
\HII\ regions with the largest equivalent widths (HCG 16 1, NGC 1533 1
\& 5). In these cases, the calculated ages are consistent with the
instantaneous burst models, and confirm that the isolated \HII\
regions are due to newly formed clusters where no stars existed
previously. A stellar age of 3$\times10^{8}$ years is found for a
continuous star formation model of NGC 1533 2. The low equivalent
width of the candidate \HII\ region in the ESO 149-G003 field however
cannot be reasonably fit by the continuous star formation models.

The calculations above assumes a simple scaling of the number of
ionizing stars with total cluster mass. At small cluster masses, the
differences between analytic and stochastic IMFs can be substantial,
especially in the number of high mass stars. Statistical errors in
stellar population models are discussed by \cite{Cervino02}. Monte
Carlo simulations can be used to investigate the effects of small
initial masses on clusters. \cite{Garcia-Vargas94} found the
probability of finding a small cluster with an $M>60$ \msun\ star is
not zero, as suggested by the analytic IMF, but rises to 12\%.  Monte
Carlo simulations by \cite{Cervino94} also obtain similar
results. These simulations suggest that the total cluster mass
calculated above could be overestimated.

\subsection{Origins of Isolated \HII\ Regions}

Since we have just two systems with confirmed isolated \HII\ regions it
is difficult to draw any conclusions on a common formation scenario,
if one exists. Evidence of interactions, however, feature in all
systems. NGC~1533 and HCG~16 both display tidally disrupted \HI\
outside the main optical region of the galaxies. Although the isolated
\HII\ region projected to be near ESO~149-G003 could be a background
object, if the two are associated, the warping in ESO~149-G003's disk
could indicate an interaction has occurred. The two systems that
feature similar objects reported in the literature, NGC 4388 in the
Virgo cluster \citep{Vollmer03, Gerhard02} and A1367 \citep{Sakai02},
also show disrupted \HI.  \cite{Vollmer03} use ram pressure stripping
to explain the isolated \HII\ region near NGC 4388. Ram pressure
stripping is less likely in our two systems, which occur in much less
dense environments.

The projected separations between the isolated \HII\ regions and host
galaxies (see Table~\ref{table:eldots1}) suggest the underlying
massive stars have mostly likely formed in situ. Alternatively, the
stars could have formed in the galaxy and then been ejected. Typical
ejection velocities due to dynamical interactions do not exceed
200$-$300 \kms\ \citep{Leonard88}. However the velocity required to
travel 4$-$33 kpc in the lifetime of a massive star ($\sim10^7$ yrs)
is 390$-$3200 \kms. The close match in velocity between the isolated
\HII\ regions and galaxy, at least in the NGC 1533 system, suggest
that the isolated \HII\ regions are not currently moving at a high
relative speed.

Our own Galactic halo has interactive \HI\ debris which appears to be
forming stars. The Magellanic Bridge is an \HI\ complex that joins the
Large and Small Magellanic Clouds at $\sim$50 kpc and represents the
interaction between these two galaxies
\citep[e.g.][]{Putman03}. Simulations indicate the Bridge was formed
200 $-$ 500 Myr ago \citep[e.g.][]{Gardiner96}, but the stars in the
Bridge are between 10 $-$ 25 Myr old \citep[e.g.][]{Demers98},
indicating that star formation is currently active within this
gas-dominated tidal feature.

We have three confirmed isolated \HII\ regions with detailed \HI\
information in the NGC~1533 system. The high velocity dispersions and
gradients in the vicinity of the isolated \HII\ regions in this system
suggest that star formation is not occurring via the usual
gravitational collapse methods.  Star formation could be shock-induced
by clouds colliding \citep[e.g.][]{Zhang01, Sato00}.  Is it reasonable
to expect collisions in the NGC 1533 system?  \cite{Christodoulou97}
calculated the timescale for collisions in low density environments
such as Galactic HVCs and the Magellanic Stream and Bridge. Following
\cite{Christodoulou97}, a cloud on a random walk with velocity
dispersion $\sigma_v$, has a characteristic time between collisions of
$\tau_1=l/\sigma_v$, where the mean free path $l=V/(N\pi R^2)$, that
is the volume (V) divided by the number of clouds (N) with cross
section $\pi R^2$. Considering all clouds in the volume, the
characteristic time between any two collision is
$\tau_c=\tau_1/N=V/(N^2\pi R^2\sigma_v)$.

The \HI\ ring around NGC 1533 appears to be clumped on scales of no
greater than the resolution of the image ($\sim 1$\arcmin),
corresponding to a radius of 3 kpc at a distance of 21 Mpc. Of course,
the gas is likely to be clumped on smaller scales too, so this radius
is an upper limit. The density, $\rho$, can be estimated from the
surface density, \nhi, where $\rho= m_H\nhi/(2R)$=
1.8$\times10^{-26}$ g cm$^{-3}$. The mass of each cloud is then given
by \mhi = 4$\pi\rho R^{3}/3$ = 3.0$\times10^{7}$ \msun. Since the
entire ring has an \HI\ mass of 7$\times10^{9}$ \msun, it could be
composed of 200 such clouds. The total ring surface area covered by
\HI\ with column densities greater than 2$\times10^{20}$ \cm\ is
4$\times10^{3}$ kpc$^{2}$, assuming a thickness of at least a cloud
diameter (6 kpc), gives a volume estimate of 24$\times10^{3}$
kpc$^{3}$. The velocity dispersion in the ring varies from 5 \kms\ in
the NW part to 30 \kms\ in the SE. The timescale of a collision
between any two clouds is therefore,
\begin{equation}
\tau_c = 6.9\times10^{5}\left(\frac{V}{24\times10^3
  \, \mathrm{kpc}^3}\right)\left(\frac{N}{200}\right)^{-2}\left(\frac{R}{3
 \, \mathrm{kpc}}\right)^{-2}\left(\frac{\sigma_v}{30\, \mathrm{kms}^{-1}}\right)^{-1} \quad \mathrm{yr}.
\end{equation}

\noindent This timescale varies between $1-4\times10^{6}$ yrs
depending on $\sigma_v$ (and is a lower limit for smaller cloud
radii). The lifetime of an O star is only 1$\times10^{7}$ yrs, so we
would expect to see $2-14$ star formation events due to cloud
collisions at any one time. This estimate agrees with the fact we see
5 isolated \HII\ regions in the NGC 1533 ring. This formation scenario
is therefore plausible.

\subsection{IGM Enrichment}

Although the \Ha\ luminosities are small, an estimate of the star
formation rate can be obtained by the relation SFR (\msun yr$^{-1}$) =
L$_{H\alpha}$/1.26$\times10^{41}$ erg s$^{-1}$
\citep{kennicutt98}. Summing the \Ha\ luminosities from the 5 isolated
\HII\ regions in the NGC~1533 system, the SFR $=1.5\times10^{-3}$ \msun
yr$^{-1}$. The other systems with only one isolated \HII\ region are
lower still. A small, but finite intergalactic star formation rate will
continually enrich and ionize the IGM. \cite{Maeder92} calculated the
total yield (y) of metals expelled in winds and ejecta from supernova
and planetary nebula. For a Salpeter IMF and a range of initial
metallicities, they find 0.022$<$y$<$0.027. Therefore a SFR
$=1.5\times10^{-3}$ \msun yr$^{-1}$ will return $\sim$4$\times10^{-5}$
\msun yr$^{-1}$ of metals into the surrounding medium. Simulations of
the dynamical evolution of \HI\ gas around NGC 1533 show that it could
last up to 1 Gyr \citep{Ryan-Weber03a}. This is considered an upper
limit since no consumption of gas due to the formation of stars in taken
into account. If the SFR is maintained for 1 Gyr, metals will pollute
the 2.4$\times10^{9}$ \msun\ of \HI\ in the SE cloud, resulting in a
metallicity of $\sim1\times10^{-3}$ solar. Alternatively if the SFR
was not continuous and corresponded to a single population of stars
only, the resulting metallicity would be negligible
($\sim1\times10^{-6}$ solar).

How does this compare to the abundances seen in \Lya\ absorption line
systems? \HI\ in the vicinity of the NGC 1533 isolated \HII\ regions
has \nhi\ $=1-4\times10^{20}$ \cm, equivalent to a damped \Lya\
absorption (DLA, \nhi\ $\geq2\times10^{20}$ \cm) or sub-DLA system
($10^{19}<$ \nhi\ $<2\times10^{20}$ \cm). The metallicity of DLA or
sub-DLA gas at low redshift varies from 0.01 solar \citep[e.g I Zw
18][]{Aloisi03} to solar. Depending on the initial metallicity, the
isolated \HII\ regions would enrich the NGC 1533 system by 10 percent
at the most. At higher redshifts however, this increase in metallicity
could be more significant. \citep{Prochaska03} find a DLA `metallicity
floor' at $\sim1.4\times10^{-3}$ solar, over a redshift range from 0.5
to 5. Intergalactic star formation may have contributed to this. DLAs
with larger velocity widths are found to have higher metallicities
\citep{Nestor03}, this trend is also hinted upon in the sub-DLA data
\citep{Peroux03}. Larger velocity widths may indicate interacting
systems. In addition, since collisions and tidal disruptions of
galaxies were more common at higher redshifts, the amount of high
\nhi-gas outside galaxies was greater and therefore the intergalactic
star formation rate could have been higher in the past.

\subsection{Kinematics and the Tidal Dwarf Galaxy Connection}

\begin{figure} 
\vspace{9cm}
\includegraphics{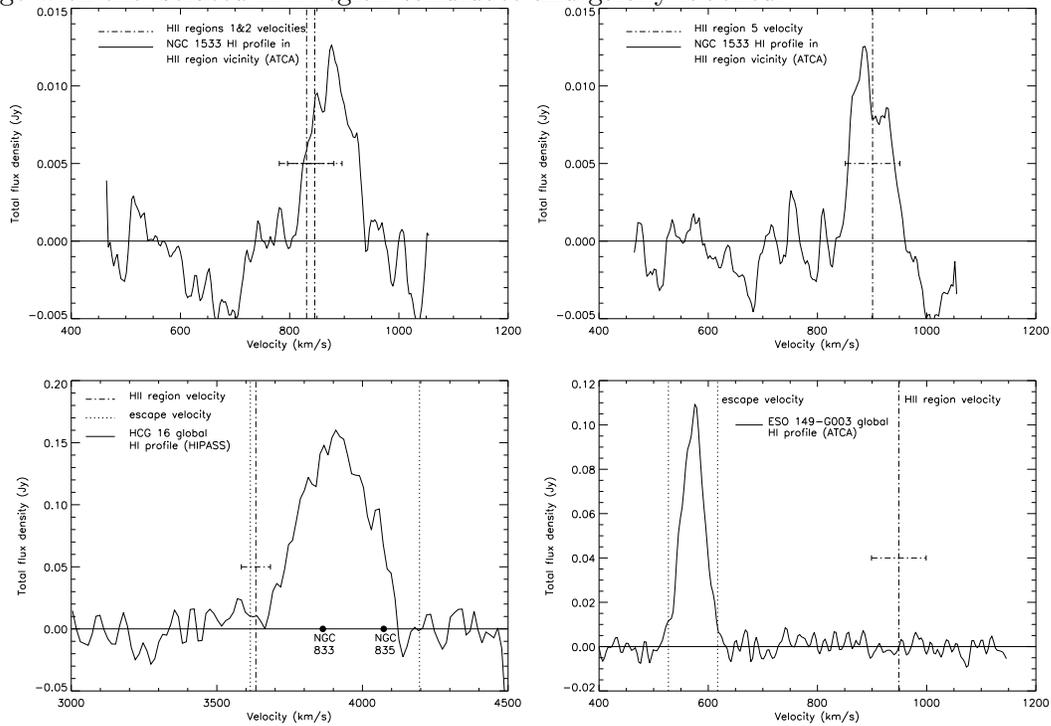} 
\caption{\HI\ spectra of the three systems with detected isolated
  \HII\ regions, the velocities of the \HII\ region candidates
  are also plotted. }
\label{fig:hispec}
\end{figure}

Comparing the velocity of isolated \HII\ regions to their apparent
host galaxy and associated \HI\ gas is useful in determining their
dynamical connection.  In Figure~\ref{fig:hispec} the \HI\ spectrum
for each system is plotted with the velocity of the detected emission
line objects, assuming the line is indeed \Ha. For the NGC 1533
system, the total \HI\ profile in a beam area centered on the isolated
\HII\ region (or regions in the case of 1 and 2, since they are so
close) is given. The velocity of all 3 NGC 1533 confirmed isolated
\HII\ regions coincide well with \HI\ gas which is bound to and
rotating around NGC 1533 \citep{Ryan-Weber03a}. Are these isolated
\HII\ regions progenitors to tidal dwarf galaxies? Since the gas and
isolated \HII\ regions are bound to the galaxy, it is likely that the
stars formed will also remain bound in the tidal debris. There is
certainly a reservoir of gas from which more stars could form, so it
is possible in this case that a tidal dwarf galaxy could emerge.

For the other two systems the global \HI\ spectrum is given, since we
do not have a synthesis map of HCG 16 and there is no \HI\ detected at
the position and velocity of the isolated \HII\ region near ESO
149-G003. The galaxy escape velocity and velocity of the isolated
\HII\ region are plotted on each spectrum to determine their kinematic
connection. The escape velocity is estimated by
$v_{esc}\sim1/\sqrt2w_{20}$sin($i$), where $w_{20}$ is the width of
the global \HI\ profile at 20\% of its height and $i$ is the
inclination of the galaxy. No inclination correction is made for HCG
16 since there is more than one galaxy embedded in the \HI\ emission.
In HCG 16 the velocity of the isolated \HII\ region sits just inside
the escape velocity of the system. It is unclear whether the stars
formed will remain bound or whether they will disperse into the
intragroup medium.  The large difference in velocity between ESO
149-G003 and its candidate isolated \HII\ region suggest the two are
not bound. Of course to form a tidal dwarf galaxy, tidal gas is
needed, the non-detection of high column density \HI\ gas in the
vicinity of this \HII\ region candidate rules out this possibility.

\subsection{ESO~149-G003: A True Isolated \HII\ Region?}
\label{sect:eso149real}

The emission line source near ESO~149-G003 could be a part of an
associated extragalactic \HI\ cloud (although technically \HI\ clouds
don't have optical counterparts). Its \HI\ mass upper limit of
$8\times10^5$ \msun\ has implications for the search for extragalactic
\HI\ clouds around other galaxies and in galaxy groups. For example,
\cite{Zwaan01} searched analogues of the Local Group to a 4.5$\sigma$
limit of $7\times10^6$ \msun\ and found no significant extragalactic
\HI\ clouds. This may motivate the search for lower \HI\ mass
extragalactic clouds. This isolated \HII\ region candidate is quite
different from the others discussed in this paper. Only one emission
line is detected (making the line identification ambiguous), the
continuum flux is significantly higher and the apparent velocity
difference between ESO~149-G003 and the \HII\ region candidate
(assuming the detected emission line is in fact \Ha, see
Figure~\ref{fig:hispec}) places the it well outside the escape
velocity of ESO~149-G003 (45 \kms). This kinematic evidence suggests
that the source is perhaps a distant emission line galaxy in the
field, rather than being associated with ESO~149-G003. Whether the
emission line is \Ha\ at 995 \kms\ or another line at a higher
redshift is uncertain.  Indeed, we expect a contamination rate of
background emission line systems of $\sim1$ per SINGG image, based on
number statistics from \cite{Boroson93}, \cite{Cowie98} and
\cite{Rhoads00}. If the emission line is \Ha, this object holds
interesting implications for the census of intergalactic
matter. Absorption lines along random lines-of-sight tell us that the
IGM is generally clumped spatially and in velocity, and shows a range
of densities and metallicities. In future studies isolated \HII\
regions could be used as beacons for star-forming regions of the
IGM. This would complement absorption studies along random sight lines
and low spatial resolution \HI\ emission observations of the IGM.

\section{Conclusion}
\label{sect:conclusion}

The discovery of intergalactic \HII\ regions presented here and in
other recent publications provides a small but finite source of
enrichment and ionization of the IGM. In two cases the fact that these
emission line objects are detected in both \Ha\ and \oiii\ rules out
the possibility that they are background emitters. The \Ha\
luminosities imply that each isolated \HII\ region is ionized by 4$-$7
O stars. If these stars have formed in situ they represent atypical
star formation in a low density environment. The low level of
continuum emission from three of four confirmed isolated \HII\ regions
suggests the stellar populations are very young and have formed where
no stars existed previously. If part of a normal IMF, the
corresponding total cluster mass would be $\sim10^{3}$ \msun. In two
out of three systems, isolated \HII\ regions are associated with tidal
\HI\ features, providing a reservoir of neutral gas. In one particular
system, NGC~1533, the mass, distribution and velocity dispersion of
the \HI\ suggests the rate of star formation ($1.5\times10^{-3}$ \msun
yr$^{-1}$) could be sustained by the collision of clouds. This would
result an increase in the metal abundance by $\sim1\times10^{-3}$
solar. This is the same abundance level as seen in the DLA
`metallicity floor' \citep{Prochaska03}. The amount of intergalactic
high column density \HI\ and rate of collision-triggered intergalactic
star formation may have been higher in the past. On-going
investigations into the metallicities and underlying stellar
population of these and other isolated \HII\ regions in the SINGG
images will shed more light on their nature and origin.

\section{Acknowledgements} 

This research has made use of the NASA/IPAC Extragalactic Database
(NED). Digitized Sky Survey (DSS) material (UKST/ROE/AAO/STScI) is
acknowledged. ERW acknowledges support from an Australia Postgraduate
Award. MEP acknowledges support by NASA through Hubble Fellowship
grant HST-HF-01132.01 awarded by the Space Telescope Science
Institute, which is operated by AURA Inc. under NASA contract NAS
5-26555. MSO acknowledges support from the National Science Foundation
Grant AST-0204853. Helpful comments by the referee are gratefully
acknowledged.

\bibliographystyle{apj}
\bibliography{eldots}

\clearpage

\clearpage

\begin{table}
\scriptsize{
\begin{tabular}{lccccccc}
Target &\multicolumn{2}{c}{Time Exposure} & Seeing &
\multicolumn{2}{c}{Detection Limit} & \multicolumn{2}{c}{Surface
  Brightness Limit} \\
Name & R & \Ha\ & & R & \Ha\ & R & \Ha\ \\
   & sec & sec & arcsec  & erg
s$^{-1}$cm$^{-2}$\AA$^{-1}$& erg s$^{-1}$cm$^{-2}$ & 
erg s$^{-1}$cm$^{-2}$\AA$^{-1}$arcsec$^{-2}$& erg s$^{-1}$cm$^{-2}$arcsec$^{-2}$ \\
\hline
HCG 16        & 360 &1800 &1.7 &1.08$\times10^{-18}$ &  1.48$\times10^{-16}$ &1.33$\times10^{-19}$ &1.51$\times10^{-17}$\\
ESO 154-G023  & 360 &1800 &1.2 &9.20$\times10^{-19}$ &  7.54$\times10^{-17}$ &5.57$\times10^{-19}$ &1.20$\times10^{-17}$\\
NGC 1314      & 360 &1800 &1.5 &9.46$\times10^{-19}$ &  1.19$\times10^{-16}$ &1.05$\times10^{-19}$ &9.29$\times10^{-18}$\\
NGC 1533      & 480 &1800 &1.4 &9.08$\times10^{-19}$ &  9.96$\times10^{-17}$ &4.99$\times10^{-19}$ &1.17$\times10^{-17}$\\
IC 5052       & 360 &1800 &1.4 &1.23$\times10^{-18}$ &  1.11$\times10^{-16}$ &2.46$\times10^{-19}$ &1.82$\times10^{-17}$\\
ESO 238-G005  & 360 &1800 &1.3 &1.01$\times10^{-18}$ &  1.04$\times10^{-16}$ &9.38$\times10^{-20}$ &1.20$\times10^{-17}$\\
ESO~149-G003  & 360 &1800 &1.7 &1.25$\times10^{-18}$ &  1.37$\times10^{-16}$ &2.72$\times10^{-19}$ &1.33$\times10^{-17}$\\
\hline           
\end{tabular}
\caption{Properties of R and \Ha\ SINGG images with isolated \HII\
  region candidates. The detection limit is the 5$\sigma$ point
  source detection limit. The final two columns give the large scale
  surface brightness limit calculated as described in the text.}
\label{table:images}
}
\end{table}

\begin{table}
\scriptsize{
\begin{tabular}{llllcccc}
Candidate \HII\ & m$_{R}$ & R Flux & H$\alpha$ Flux & EW
&H$\alpha$ Lum.&Lines&Velocity\\ 
Region Name&AB mag & erg s$^{-1}$cm$^{-2}$\AA$^{-1}$&erg
s$^{-1}$cm$^{-2}$ & \AA& erg s$^{-1}$& Detected&\kms\\
\hline
HCG~16 1      &$>23.5$	     &  $<1.08\times10^{-18}$          &1.0($\pm$0.05)$\times10^{-15}$ & $>$1764  & 3.5$\times10^{38}$ &\Ha, \oiii &3634($\pm$50)\\ 
ESO 154-G023 1&$21.3\pm0.1$   &	7.6($\pm0.5)$$\times10^{-18}$  &3.8($\pm$0.5)$\times10^{-16}$  & 51       & 1.3$\times10^{36}$&none&...\\
NGC 1314 1    &$21.9\pm0.1$   &	4.6($\pm0.5)$$\times10^{-18}$  &1.6($\pm$0.05)$\times10^{-15}$ & 431      & 5.2$\times10^{38}$&none&...\\
NGC~1533 1    &$23.1\pm0.3$   &	1.5($\pm0.4)$$\times10^{-18}$  &1.1($\pm$0.5)$\times10^{-15}$  & 1113     & 5.8$\times10^{37}$&\Ha, \oiii& 846($\pm$50)\\ 
NGC~1533 2    &$22.0\pm0.1$   &	4.1($\pm0.4)$$\times10^{-18}$  &6.8($\pm$0.5)$\times10^{-16}$  & 177      & 3.6$\times10^{37}$&\Ha&831($\pm$50) \\
NGC~1533 3    &$22.6\pm0.2$   &	2.3($\pm0.4)$$\times10^{-18}$  &6.4($\pm$0.5)$\times10^{-16}$  & 331      & 3.4$\times10^{37}$&not obs. &...\\
NGC~1533 4    &$22.5\pm0.2$   &	2.5($\pm0.4)$$\times10^{-18}$  &5.8($\pm$0.5)$\times10^{-16}$  & 262      & 3.1$\times10^{37}$&not obs. &...\\
NGC~1533 5    &$>23.6$	     &  $<9.08\times10^{-19}$          &5.0($\pm$0.5)$\times10^{-16}$  & $>$1658  & 2.6$\times10^{37}$&\Ha&901($\pm$50)\\
IC 5052 1     &$22.6\pm0.3$   &	2.4($\pm0.5)$$\times10^{-18}$  &3.2($\pm$0.4)$\times10^{-16}$  & 143      & 1.3$\times10^{36}$&none&... \\
IC 5052 2     &$20.72\pm0.05$ &  1.34($\pm0.05)$$\times10^{-17}$&4.0($\pm$0.4)$\times10^{-16}$  & 31       & 1.7$\times10^{36}$&none&...  \\
ESO 238-G005 1&$23.1\pm0.3$   &	1.5($\pm0.5)$$\times10^{-18}$  &1.4($\pm$0.6)$\times10^{-16}$  & 102      & 1.3$\times10^{36}$&none&... \\
ESO 238-G005 2&$>23.5$	     &  $<1.3\times10^{-19}$           &2.0($\pm$0.6)$\times10^{-16}$  & $>$1778  & 1.9$\times10^{36}$&none&... \\
ESO~149-G003 1&$20.39\pm0.03$ &  1.8($\pm0.05)$$\times10^{-17}$ &6.9($\pm$0.5)$\times10^{-16}$  & 39       & 3.5$\times10^{36}$&\Ha? &949($\pm$50)\\

\hline           
\end{tabular}
\caption{Properties of isolated \HII\ region candidates with DBS
  spectra. Spectroscopically detected isolated \HII\ regions have
  velocities listed in the final column. The fluxes and equivalent
  widths are measured from the SINGG images. In three cases, where the
  continuum flux is below the detection limit, the upper limit EW is
  given. The \HII\ regions candidates 3 \& 4 of NGC 1533 are included
  in this and the following table even though no spectra were taken.}
\label{table:eldots2}
}
\end{table}

\begin{table}
\begin{minipage}{160mm}
\scriptsize{
\begin{tabular}{lccccccc}
Host&Velocity&Distance&\HII\ Reg. & Separation 
&\multicolumn{2}{c}{\HII\ Reg. Position}& \nhi\ \\   
Galaxy&\kms&Mpc&No. &kpc, R$_{25}$&RA &Dec & cm$^{-2}$ \\
\hline
HCG~16&     3917&53 & 1&19, 2.0&02:09:28&  -10:07:16   & ...                  \\
ESO 154-G023&574&5.3& 1&6.0, 3.5&02:56:31&  -54:31:35  & 2.8$\times10^{19}$   \\
NGC 1314    &3936&52& 1&131, 15.2&03:23:12& -04:15:16  & ...                  \\
NGC~1533    &785&21 & 1&33, 4.0&04:10:13&  -56:11:37   & 2.4$\times10^{20}$   \\
NGC~1533    &785&21 & 2&33, 4.0&04:10:14&  -56:11:35   & 2.4$\times10^{20}$   \\
NGC~1533    &785&21 & 3&31, 3.8&04:10:17&  -56:10:46   & 9.8$\times10^{19}$   \\
NGC~1533    &785&21 & 4&16, 2.0&04:10:11&  -56:07:28   & 3.2$\times10^{20}$   \\
NGC~1533    &785&21 & 5&20, 2.5&04:10:15&  -56:06:15   & 1.5$\times10^{20}$   \\
IC 5052     &584&5.9& 1&8.0, 1.4&20:52:59&  -69:12:27  &$<$3.2$\times10^{19}$ \\
IC 5052     &584&5.9& 2&10, 1.7&20:52:53&  -69:16:22   &$<$3.2$\times10^{19}$ \\
ESO 238-G005&706&8.9&1 &3.9, 30&22:22:33&  -48:25:42   & 2.4$\times10^{20}$   \\
ESO 238-G005&706&8.9&2 &11, 84&22:22:42&  -48:27:53    &$<$3.2$\times10^{19}$ \\
ESO~149-G003&576(628)\footnote{Velocity of the \Ha\ line in the galaxy
  measured from the same long-slit spectrum as the isolated \HII\ region is 628 ($\pm$50)
  \kms.}&6.5&1&3.4, 5.2&23:51:51&-52:34:34 &$<$3.2$\times10^{19}$\\   
\hline           
\end{tabular}
\caption{Isolated \HII\ region candidates with DBS spectra. The
  velocity is the heliocentric velocity of the host galaxy measured by
  HIPASS. The distance given is to the host galaxy, HCG 16 uses the
  mean recessional velocity from \cite{Ribeiro98}, NGC 1533 uses the
  distance from \cite{Tonry01} and all other galaxies use the local
  group corrected velocity. The projected separation in kpc (to the
  optical centre) is calculated using this distance. The projected
  separation is also calculated as a fraction of R$_{25}$, the major
  axis of the isophote at $\mu_R = 25 {\rm mag\, arcsec^{-2}}$. The
  \HI\ column densities are measured from ATCA maps, at the position
  of the \HII\ region candidates, where available.}
\label{table:eldots1}
}
\end{minipage}
\end{table}

\end{document}